\documentclass[preprint,showpacs,preprintnumbers,amsmath,amssymb]{revtex4}

\usepackage{graphicx}% Include figure files
\usepackage{dcolumn}% Align table columns on decimal point
\usepackage{bm}% bold math

\begin{document}

%\preprint{FT-HEP022}

\title{Role of the $\rho(1450)$ in low-energy observables from an analysis in the meson dominance approach.}% Force line breaks with \\

\author{Gustavo \'Avalos}
\affiliation{Instituto de F\'{\i}sica,  Universidad Nacional Aut\'onoma de M\'exico, AP 20-364,  M\'exico D.F. 01000, M\'exico}%
\author{Antonio Rojas}
\affiliation{Instituto de F\'{\i}sica,  Universidad Nacional Aut\'onoma de M\'exico, AP 20-364,  M\'exico D.F. 01000, M\'exico}%
\author{Marxil S\'anchez}
\affiliation{Instituto de F\'{\i}sica,  Universidad Nacional Aut\'onoma de M\'exico, AP 20-364,  M\'exico D.F. 01000, M\'exico}%
\author{Genaro Toledo}
\affiliation{Instituto de F\'{\i}sica,  Universidad Nacional Aut\'onoma de M\'exico, AP 20-364,  M\'exico D.F. 01000, M\'exico}%

\date{\today}% It is always \today, today,
\begin{abstract}
The $\rho(1450)$ vector meson ($\rho^\prime$) is becoming increasingly important to properly describe precision observables. We analyse a set of decay modes and cross sections, in the low-energy regime, to determine the role played by this meson. This is done through the extraction of the parameters for its description, in the context of the vector meson dominance model and its effective hadronic interactions, involving the low mass lying hadrons ($\rho$, $\omega$ and $\pi$). In a first step, we determine the parameters of the model from ten decay modes which are insensitive to the $\rho^\prime$. Then, we consider the $\omega \rightarrow 3\pi$ decay and exhibit the need to extend the description, by incorporating the $\rho^\prime$ and a contact term as prescribed by the Wess-Zumino-Witten anomaly. In a second step, we incorporate the data from the $e^+e^- \rightarrow 3\pi$ cross section (as measured by SND, CMD2, BABAR and BESIII) and then the $e^+e^- \rightarrow \pi^0 \pi^0 \gamma$ data (as measured by SND and CDM2) to further restrict the $\rho^\prime$ parameters validity region. As an application of the results, we compute the $e^+e^- \rightarrow 4\pi$ cross section for the so-called omega channel, measured by BABAR, and find a good description of the data considering the parameters found. As a byproduct, the coupling between $\rho$, $\omega$ and $\pi$ ($g_{\rho\omega\pi}= 11.314 \pm 0.383$ GeV$^{-1}$) is found to be consistent with all the relevant observables, upon the inclusion of the $\rho^\prime$ and the contact term.

\end{abstract}

%\pacs{13.40.Gp, 11.10.St, 14.40.-n}% PACS, the Physics and Astronomy
\maketitle

\section{Introduction}

The low-energy measurements involving hadrons are reaching a high accuracy. In general, the low mass hadron spectra contributing to the processes can be identified, and the corresponding mass and decay width parameters can be obtained. Excited states may manifest themselves in low-energy observables, when not considered explicitly, as a modification of the effective interaction between nonexcited states. For energies reaching the threshold of their nominal mass, they exhibit their resonant features in the observables and are necessarily included to properly describe the data. The $\rho(1450)$ vector meson (denoted by $\rho^\prime$ wherever possible) is one example of such states. Its contribution to the $\omega \to 3 \pi$ decay width can be identified by noticing that it cannot be reproduced by only considering the $\rho$ meson as the intermediate state \cite{vmdrelation,david}. 
On the other hand, the spectra obtained in hadronic $\tau$ decays \cite{CLEO:1999heg,Belle:2008xpe} and $e^+e^-$ anihilation into hadrons \cite{snd2pg16} exhibit clear indications of its presence and are used to determine its mass and total decay width \cite{pdg}. This important information needs to be complemented with the partial width of the different decay modes, which then have implications on the parameters of the models attempting to describe them. This information has not been settled, although evidence can be extracted from particular observables \cite{pdg}.  
Decay modes such as $\rho^\prime \to \omega\pi$ and $\rho^\prime \to \pi\pi$ are of particular interest to disentangle the contribution of the $\rho^\prime$ and $\rho$ mesons in low-energy observables sensitive to both mesons. 
A combined analysis of low-energy observables show up as a possibility to provide further information on this issue.
They are involved, for example, in the $e^+e^- \to \pi^0\pi^+\pi^-$ \cite{snd3p,CMD23p,BABAR3p,BES3p} and $e^+e^- \to \pi^0\pi^0 \gamma$ \cite{snd2pg00,snd2pg13,snd2pg16,cmd22pg} processes and in $e^+e^- \to \pi^0\pi^0 \pi^+\pi^-$ process driven by the $\omega$ meson as intermediate state\cite{babar4p,snd4p}.\\

 In this work, we analyse a set of decay modes and cross sections involving the $\rho$, $\omega$ and $\rho^\prime$  mesons, described in the context of the vector meson dominance model and their effective hadronic interactions including the pion, to determine the corresponding parameters.
We perform a fit to the data making use of MINUIT package for minimization and VEGAS \cite{vegas} subroutine for the phase space integration, to obtain the cross section whenever needed. In a first step, we determine the parameters of the model involving the $\rho$, $\omega$ and $\pi$ mesons, from ten decay modes which are practically insensitive to the $\rho^\prime$, namely, $\rho \to \pi\pi$ neutral and charged modes, $\rho^0 \to e^+e^-, \ \mu^+\mu^-$, $\omega \to e^+e^-, \ \mu^+\mu^-$, $\omega \to \pi^0\gamma$, $\rho \to \pi\gamma$ neutral and charged modes and $\pi^0 \to \gamma\gamma$.
Then, we include the $\omega \rightarrow 3\pi$ decay, initially considered as driven only by the $\rho$ meson intermediate state, to exhibit the modification of the parameters previously obtained, signaling the inconsistency and therefore the need to extend the description by incorporating the $\rho^\prime$ and a contact term as prescribed by the Wess-Zumino-Witten anomaly (WZW) \cite{anomaly1,anomaly2}. In a second step, we incorporate the data from the $e^+e^- \rightarrow 3\pi$ cross section as measured by SND, CMD2, BABAR and BESIII \cite{snd3p,CMD23p,BABAR3p,BES3p}) and then $e^+e^- \rightarrow \pi^0 \pi^0 \gamma$ data as measured by SND and CDM2 \cite{snd2pg00,snd2pg13,snd2pg16,cmd22pg} to further restrict the $\rho^\prime$ parameters validity region.
As an application of the results, we compute the $e^+e^- \rightarrow 4\pi$ cross section for the so-called omega channel, and compare with the data reported by BABAR \cite{babar4p} considering the parameters previously found. As a byproduct, we keep track of the behaviour of the coupling between the $\rho$, $\omega$ and $\pi$ mesons to determine its stability, upon the inclusion of the $\rho^\prime$ and contact term in the description of the processes under consideration.

\section{Theoretical framework}
The vector meson dominance model (VMD) considers that neutral vector mesons couple to the electromagnetic current \cite{sakurai}. By considering the hadrons as the relevant degrees of freedom and including the interaction with pseudoscalar mesons, this description is able to account for the low energy manifestation of the strong interaction. Incorporation of symmetries such as isospin and $SU(3)$ flavour symmetry allow us to both classify the hadrons and relate their properties. Further considerations associated to the vector mesons manifestation as gauge bosons and incorporation of higher symmetries have been also considered as extensions of the VMD \cite{klz,vmdbando,fujiwara,vmdx}. Here, since the hadrons involved are the lightest ones, we restrict ourselves to the part that is common to all the VMD based models. 
The effective Lagrangian including the light mesons $\rho$, $\pi$ and $\omega$, in addition to the $\rho^\prime$ can be set as
\begin{eqnarray}
{\cal L}&=& \sum_{V=\rho,\,\rho^\prime} g_{V\pi\pi}\,
\epsilon_{abc}\, V_\mu^a\, \pi^b \,\partial^\mu\, \pi^c 
+\sum_{V=\rho,\,\rho^\prime} 
g_{\omega V\pi}\,\delta_{ab}\,\epsilon^{\mu\nu\lambda\sigma}\,\partial_\mu\, \omega_\nu\, \partial_\lambda\, V_\sigma^a\,  \pi^b \nonumber \\
&+&g_{3\pi}\,\epsilon_{abc}\,   \epsilon^{\mu\nu\lambda\sigma}\,\omega_\mu\, \partial_\nu\, \pi^a\,  \partial_\lambda\,  \pi^b\,  \partial_\sigma\, \pi^c +
\sum_{V=\rho,\,\rho^\prime,\,\omega} \frac{e\, m_V^2}{g_V}\,V_\mu\, A^\mu.
\label{Lvmd}
\end{eqnarray}
We have labeled the couplings with the corresponding interacting fields, and $g_{3\pi}$ is the WZW contact term. In general, $V$ refers to a vector meson, and $A^\mu$ refers to the photon field. The couplings are free parameters to be determined from experiment. Although, as we mention before, relations between them and even from other descriptions can be drawn \cite{fujiwara,kura,kay,kura1,xral,prades94}.
The mesons involved in this description are produced in experiments devoted to the hadronic production from electron-positron annihilation, as mentioned above, and hadronic tau decays \cite{tau,kloe,aulchenko,achasov02,achasov03,pdg}. Here, we consider only data from the former.
The strong interaction between the $\omega$, $\rho$ and $\pi$ mesons, encoded in the $g_{\omega \rho \pi}$ parameter, necessarily involves at least one of the particles off-shell due to phase space restrictions.
The $g_{\omega \rho \pi}$ coupling, the $\rho^\prime$  parameters and the $\omega \rightarrow 3\pi$ contact term ($g_{3\pi}$) usually appear together when describing experimental data, exhibiting a strong correlation among them \cite{david}. Therefore, an analysis involving data from different sources should help to disentangle their individual contributions. This information is relevant in the understanding of other scenarios, where there is not enough information to draw an independent analysis, and therefore it requires us to rely on a well-supported determination of such parameters to draw conclusions.\\
In the following, we describe the generic processes and the way they are incorporated in the analysis. We will extend our discussion on each contribution and the works related to them along the text. 

\section{ $V \to P_{1}\,P_{2}$ decay and the $g_{VP_{1}P_{2}}$ coupling}
The coupling of a vector meson ($V$) and two pseudoscalar mesons ($P$), denoted in general by $g_{VP_{1}P_{2}}$, can be extracted from the measurement of the $V\to P_{1}\,P_{2}$ decay width. The amplitude of this process, depicted in Fig. \ref{fig1}(a), can be written as: 
\begin{eqnarray}
\mathcal{M}=i\,g_{VP_{1}P_{2}}\,(p_{1}-p_{2})^{\mu}\,\eta_{\mu}(q),
\end{eqnarray}
where $q$, $p_{1}$ and $p_{2}$ are the momenta of the initial vector meson $V$ and the pseudoscalar pair in the final state, respectively. $\eta_\mu$ is the polarization tensor of the vector particle. 
The decay width is given in terms of the coupling and the masses of the particles involved as
\begin{eqnarray}
\Gamma_{VP_{1}P_{2}} =\frac{g_{VP_{1}P_{2}}^2\,\lambda^{3/2}(m^{2}_{V},m^{2}_{P_{1}},m^{2}_{P_{2}})}{48\,\pi\, m^{5}_{V}},
\label{vppwidth}
\end{eqnarray}
where $m_{V}$, $ m_{P_{1}}$ and $m_{P_{2}}$ are the corresponding masses and
$\lambda (x,y,z) =x^2+y^2+z^2-2xy-2xz-2yz $ is the K\"allen function.
As we can see in the Eq. (\ref{vppwidth}), the $g_{VP_{1}P_{2}}$ coupling is dimensionless. This result can be applied, for example, to obtain both $g_{\rho \pi \pi}$ and $g_{\rho^\prime \pi \pi}$, provided the data for the partial decay width are available. This is the case for the $\rho \to \pi\,\pi$ decay \cite{pdg}. In Table \ref{tablegvpp}, we show the values of $g_{\rho \pi \pi}$ from two different process, $\rho^{0}(770)\to \pi^{+}\,\pi^{-}$ and $\rho^{+}(770)\to \pi^{+}\,\pi^{0}$ and their weighted average. The weighted average and its uncertainty are defined in general as (see the review on statistics in Ref.\cite{pdg}), 
\begin{eqnarray}
\bar{x}\pm \delta\bar{x}= \frac{\sum^{n}_{i}\,w_{i}\,x_{i}}{\sum^{n}_{i}\,w_{i}} \pm (\sum^{n}_{i}\,w_{i})^{-1/2},
\end{eqnarray}
where $x_{i}$ and $\delta\bar{x}$ are the $i$ value and error of the $i$ measurement and $w_{i}=1/(\delta{x}_i)^2$ is the $i$ weight associated with this measurement. In our case, $x_{i}$ refers to the coupling constant.\\ 
For the $\rho^\prime$, the partial decay width is not settled; thus, we can use Eq.~(\ref{vppwidth}) and its experimental total width of 400 MeV to set the upper bound $g_{\rho^\prime \pi \pi} \leq 6.64$, considering this to be the only decay mode. We will enforce this restriction to set the region for this parameter, as we discuss later. The $\rho^\prime \to \pi\,\pi$ decay can be also addressed in an indirect way, by considering it as part of a decay chain. For example, in the $D_s \to \rho(1450)\,\pi$ decay, where the $\rho(1450)$ is reconstructed using the two pions decay mode, albeit of requiring additional information on other couplings \cite{marxilDs}.

\begin{table}[htb]
\begin{center}
\begin{tabular}{lcc}
\hline
\hline
 Process &\hspace{1 cm} &$g_{\rho\pi \pi}$ \\
\hline
 $\rho^{0} (770)\rightarrow \pi^{+} \pi^{-}$ & & $5.944\pm 0.018$\\

 $\rho^{+} (770)\to \pi^{+} \pi^{0}$& &$5.978\pm 0.048$\\

Weighted Average & &$5.953\pm 0.017$\\
\hline
\hline
\end{tabular}
\end{center}
\caption{ $g_{\rho\pi \pi}$ coupling from the neutral and charged processes and the weighted average, $\overline{g}_{\rho \pi\pi}$.}
\label{tablegvpp}
\end{table}

\section{ $V \to l\,l$ decay and the $g_{V}$ coupling}
\begin{figure}[htb]
\begin{center}
\includegraphics[scale=0.7]{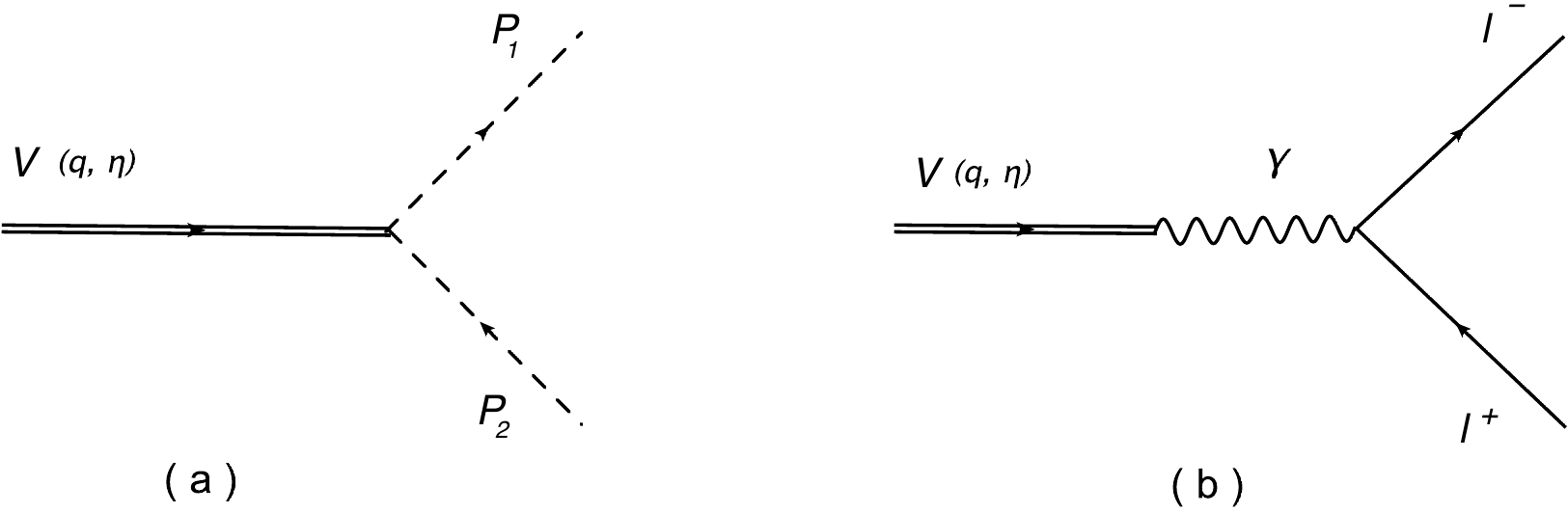}
\end{center}
\caption{Decay of vector mesons of the form (a)$V \to P\,P$ and (b)$V\to l\,l$.}
\label{fig1}
\end{figure}

The vector-photon transition depends on the $g_{V}$ coupling, as given in Eq. (\ref{Lvmd}). It can be extracted from the measurement of the $V\to \ell^{+}\,\ell^{-}$ decay width, with $\ell$ being either electrons or muons. The amplitude of this process, depicted in Fig. \ref{fig1}(b), can be written as 
\begin{eqnarray}
\mathcal{M}=-i\,\frac{e^{2}}{g_{V}}\,\bar{u}(l_{1})\,\gamma^{\nu}\,v(l_{2})\,\eta_{\nu}(q),
\end{eqnarray}
where $q$, $l_{1}$ and $l_{2}$ are the momenta of the initial vector meson and the lepton pair in the final state, respectively. $\eta_{\nu}$ is the polarization tensor of $V$ and $\bar{u}(l_{1})$ and $v(l_{2})$ are the corresponding spinors of the leptons. Then, the decay width, $\Gamma_{V \ell \ell}$, is given in terms of the coupling, the mass of the vector meson $m_{V}$ and the mass of the leptons $m_{\ell}$ as:
\begin{eqnarray}
\Gamma_{V\ell \ell} =\frac{4\,\pi\,\alpha^{2} (2\,m^{2}_{\ell}+m^{2}_{V})\,(m^{2}_{V}-4\,m^{2}_{\ell})^{1/2}}{ 3\,m^{2}_{V}\,g_{V}^2}. 
\label{vllwidth}
\end{eqnarray}
Notice that $g_{V}$ is dimensionless, as we can see in the equation above.  
In Table \ref{tablevll}, we show the values of the $g_{V}$ couplings for a set of vector mesons obtained from decays to muon and/or electron pairs. Note that we have included the value for $g_{\rho}(1450)$ obtained from the information in  Ref.\cite{pdg} but quote only a central value, as the experimental information provides only an estimate of the decay width. Improvements on this measurement would be very useful. Still, it will help us to guide the analysis on this parameter when considering scattering processes. The weighted average $\overline{g}_{V}$ from the $V\to \mu^{+}\,\mu^{-} $ and $V\to e^{+}\,e^{-}$ decays is shown in Table \ref{tablevlla} for $\rho(770)$, $\omega(782)$ and $\phi(1020)$ mesons.

\begin{table}[htb]
\begin{center}
\begin{tabular}{lcccc}
\hline
\hline
 Process &\hspace{1 cm }& Coupling &\hspace{1 cm}& Value\\
\hline
$\rho^{0}(770)\to e^{+}\,e^{-}$& &$g_{\rho}$ & & $4$.$956\pm 0.021$\\

$\rho^{0}(770)\to \mu^{+}\,\mu^{-}$& &$g_{\rho}$ & &$5$.$037\pm 0.021$\\

 $\omega(782)\to e^{+}\,e^{-}$& &$g_{\omega}$ & & $17$.$058\pm 0.292$\\

$\omega(782)\to \mu^{+}\,\mu^{-}$& &$g_{\omega}$ & &$16$.$470\pm 2.469$\\

$\phi(1020) \to e^{+}\,e^{-} $& & $g_{\phi}$ & & $13$.$381\pm 0.216$\\

$\phi(1020) \to \mu^{+}\,\mu^{-} $& & $g_{\phi}$ & & $13$.$674\pm 0.479$\\

$\rho(1450)\to e^{+}\,e^{-}$& & $g_{\rho(1450)}$ &  &$13.528$\\
\hline
\hline
\end{tabular}
\end{center}
\caption{ $g_{V}$ ($V=\rho (770)$, $\omega (782)$, $\phi (1020)$, $\rho (1450)$) couplings from decays to muon and/or electron pairs. For the $g_{\rho(1450)}$ we quote only a central value, as the experimental information provides only an estimate of the decay width.}
\label{tablevll}
\end{table}

\begin{table}[htb]
\begin{center}
\begin{tabular}{lcc}
\hline
\hline
 Coupling & &Value\\
\hline
 $\overline{g}_{\rho}$ & &$4$.$966\pm 0$.021\\

$\overline{g}_{\omega}$ & &$16$.$972\pm 0$.287\\

$\overline{g}_{\phi}$ & &$13$.$528\pm 0$.339\\
\hline
\hline
\end{tabular}
\end{center}
\caption{Weighted average couplings $\overline{g}_{V}$ ($V=\rho$, $\omega$, $\phi$).}
\label{tablevlla}
\end{table}

\section{$V_{1} \to P\,\gamma$ decay and the $g_{V_{1}P\gamma}$ and $g_{V_{1}V_{2}P}$ couplings}

\begin{figure}
\begin{center}
\includegraphics[scale=0.7]{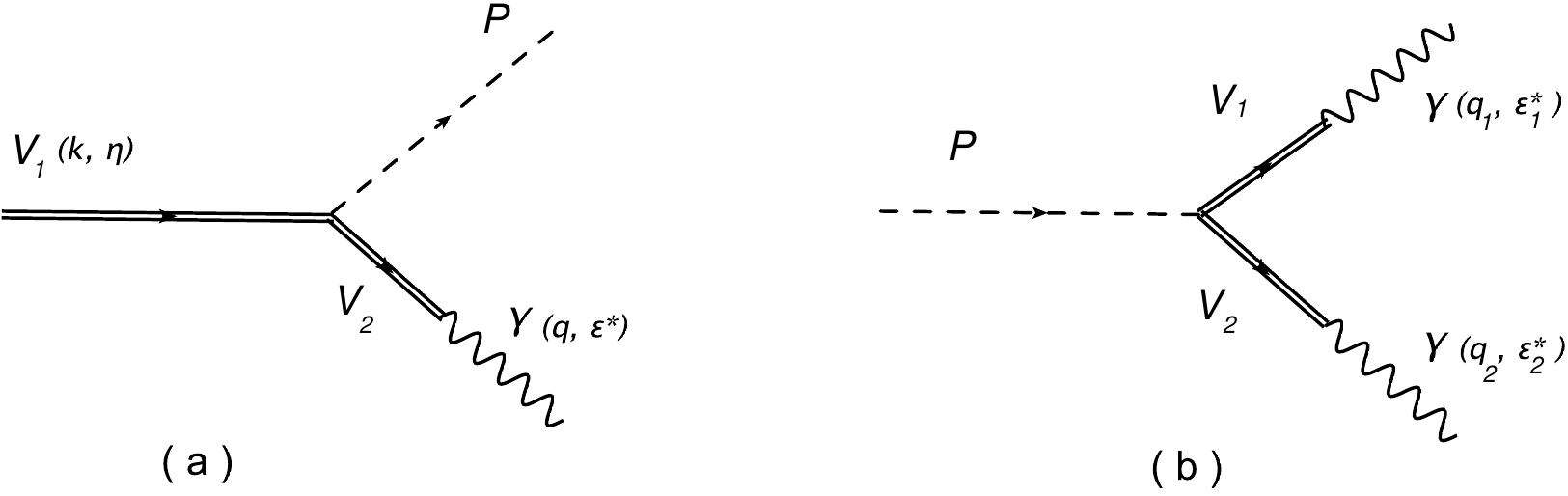}
\end{center}
\caption{Decay of vector mesons of the form (a)$V \to P\,\gamma$ and (b)$P\to \gamma\,\gamma$.}
\label{vradfig}
\end{figure}

The $g_{V_{1}P\gamma}$ coupling can be extracted from the $V_{1}\to P\,\gamma$ decay width, where $V_{1}$ is a vector meson, $P$ is a pseudoscalar meson, and $\gamma$ is the photon. The amplitude of this process, depicted in Fig. \ref{vradfig}(a), can be written as 
\begin{eqnarray}
\mathcal{M}=i\,g_{V_{1}P\gamma}\,\epsilon^{\beta \nu \alpha \mu}\, k_{\beta}\,q_{\alpha}\,\eta_{\mu}\,\epsilon^{\ast}_{\nu}
\label{vpgammam}
\end{eqnarray}
where $k$ ($\eta$) and $q$ ($\epsilon^{\ast}$) are the momenta (polarization tensor) of the $V_{1}$ and $\gamma$, respectively. 
The decay width $\Gamma_{V_{1} P\gamma}$ is given in terms of the $g_{V_{1}P\gamma}$ coupling, the masses of the vector meson $m_{V}$, and pseudoscalar meson $m_{P}$ as
\begin{eqnarray}
\Gamma_{V_{1}P\gamma}= g_{V_{1}P\gamma}^2\,\bigg [ \frac{(m_{V_{1}}^{2}-m^{2}_{P})^{3}}
{96\,\pi\, m_{P}^{3}}  \bigg ].
\label{vpgammawidth}
\end{eqnarray}

As we can notice from Eq. (\ref{vpgammawidth}), $g_{V_{1}P\gamma}$ has inverse energy units.
Another related coupling is the one where two vector mesons interact with a pseudoscalar meson, denoted by $g_{V_{1}V_{2}P}$. It can be obtained from  the previous vector meson radiative decay ($V_{1}\to P\,\gamma$) considering that the photon emission is mediated by a neutral vector meson \cite{vmdrelation}  ( See Fig. \ref{vradfig}(a)). Then, the amplitude and decay width of this process are similar to the previous Eqs. (\ref{vpgammam}) and (\ref{vpgammawidth}), with the replacement $g_{V_{1}P\gamma} \to g_{V_{1}V_{2}P}\,(e/g_{V_2})$. It follows that $g_{V_{1}V_{2}P}$ also has inverse energy units, and we use this parameter hereafter. For the analysis, we consider the following decays, with their respective charge combinations: the $\omega \rightarrow \pi\,\gamma $ decay, driven by the $\omega \rightarrow \pi\,\rho \rightarrow \pi\,\gamma $ process, and the $\rho \rightarrow \pi\,\gamma $ decay, driven by the $\rho \rightarrow \pi\, \omega \,(\phi) \rightarrow \pi\,\gamma $ processes. The contribution from the $\phi$ meson channel is relatively small and neglected at this stage \cite{lichard94, lichard11} (the expected ratios for $g_{\rho \omega\pi }/g_\omega \approx 0.7$\, can be compared with $ g_{\phi\rho\pi}/g_{\phi}\approx 0.06$, taking $g_\phi$ weighted average and, as an approach, $|g_{\phi\rho\pi}|=0.86 \pm 0.01$ GeV$^{-1}$ obtained by considering the decay width of the $\phi \rightarrow 3\,\pi$ to be fully accounted for by the $\rho\,\pi$ channel); contributions from other channels are relatively smaller \cite{kloe}. Note that the $\phi$ meson parameters are not considered as part of the analysis.
 
 \section{ $\pi^0 \rightarrow \gamma\,\gamma$ decay and the $g_{P\gamma \gamma}$ and $g_{V_{1}V_{2}P}$ couplings}
 The $g_{P\gamma \gamma}$ coupling can be extracted from the measurement of the $P\to \gamma\,\gamma$ decay width. The amplitude of this process, depicted in Fig. \ref{vradfig}(b), can be written as 
\begin{eqnarray}
\mathcal{M}=i\,g_{P\gamma\gamma}\,\epsilon^{\alpha \mu \beta \nu}\,q_{1\beta}\,q_{2\alpha}\,\epsilon^{\ast}_{1\mu}\,\epsilon^{\ast}_{2\nu}
\label{pggm}
\end{eqnarray}
where $q_{1}$ ($\eta_{1}^{\ast}$) and $q_{2}$ ($\eta^{\ast}_{2}$) are the momenta (polarization tensors) of the final photons, respectively.
We can write the decay width $\Gamma_{ P\gamma\gamma}$ in terms of the $g_{P\gamma\gamma}$ coupling and the mass of the pseudoscalar meson $m_{P}$ as
\begin{eqnarray}
\Gamma_{P\gamma\gamma}=\bigg [ \frac{g^2_{P\gamma\gamma}\,m_{P}^{3}}{64\,\pi} \bigg ].
\label{pggwidth}
\end{eqnarray}
As we can notice from the above equation, $g_{P\gamma\gamma}$ has inverse energy units.
The $g_{V_{1}V_{2}P}$ coupling can be related to this decay considering that the photons emission is mediated by two neutral vector mesons, $\pi^0 \rightarrow \rho\,\omega\,(\phi)\rightarrow \gamma\,\gamma$ \cite{vmdrelation}.
Then, the amplitude and decay width of this process are similar to Eqs. (\ref{pggm}) and (\ref{pggwidth}), by replacing 
$g_{P\gamma\gamma} \to g_{V_{1}V_{2}P}\,\frac{4\,\pi\,\alpha}{g_{V_1}\,g_{V_2} }$.
In Table \ref{tablepgg}, we show the values of the $g_{\rho \omega \pi}$ coupling from four different decays: $\omega(782)\to \pi^{0}\,\gamma$, $\rho^{0}(770)\to \pi^{0}\,\gamma$, $\rho^{+}(770)\to \pi^{+}\,\gamma$, and $\pi^{0}\to \gamma\,\gamma$. We have used the values of the $g_V$ couplings as listed in Table \ref{tablevlla} and neglected the $\rho-\phi$ channel in the $\pi^{0}\to \gamma\,\gamma$ decay.
Mixing effects from $\pi^0-\eta-\eta^\prime$ states are not considered, although they may become relevant in precision observables analysis \cite{bramon,escribano,Roig:2014uja,mateu,davier}.
 
\begin{table}[htb]
\begin{center}
\begin{tabular}{lcc}
\hline
\hline
 Process & & $g_{\rho\omega\pi}$ (GeV$^{-1}$) \\

$\omega(782)\to \pi^{0}\,\gamma$ & &$11.489\pm 0.387$ \\

$\rho^{0}(770)\to \pi^{0}\,\gamma$ & &$14.224\pm 2.227$ \\

$\rho^{+}(770)\to \pi^{+}\,\gamma$& &$12.358\pm 1.806$ \\

$\pi^{0}\to \gamma\,\gamma$&  & $11.712\pm 1.397$ \\
\hline
\hline
\end{tabular}
\end{center}
\caption{ $g_{\rho \omega \pi}$ coupling obtained from four different processes.}
\label{tablepgg}
\end{table}

\section{$\omega \to 3\,\pi$ decay}
\begin{figure}[htb]
\begin{center}
\includegraphics[scale=0.75]{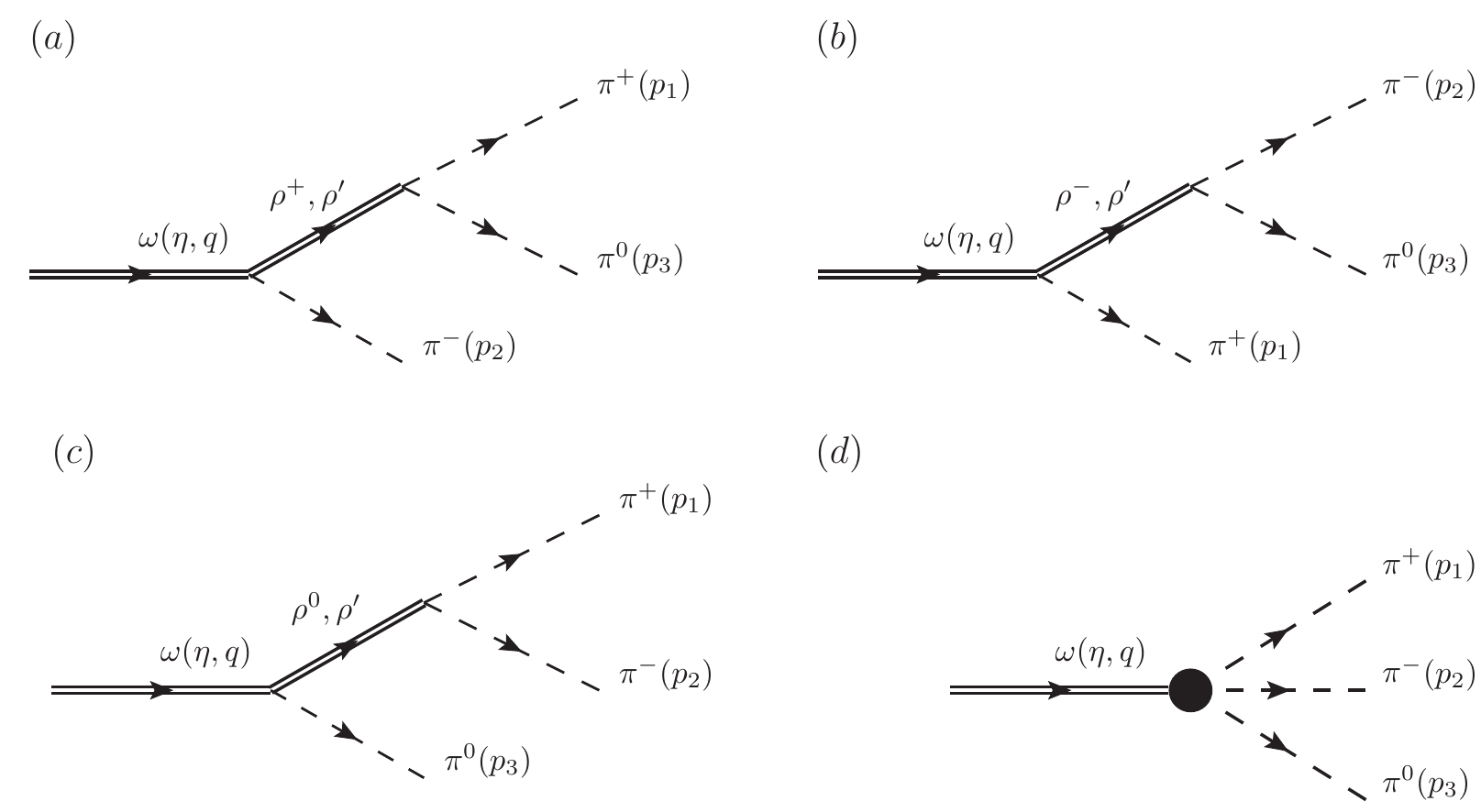}
\end{center}
\caption{Feynman diagrams for $\omega \rightarrow 3\pi$ process. The contribution from the $\rho$, $\rho^\prime$ ($a,b,c$) and the contact ($d$) channels.}
\label{ome_3pifig}
\end{figure}

Let us consider the decay process $\omega(\eta, q)\rightarrow \pi^{+}(p_1)\,\pi^{-}(p_2)\,\pi^{0}(p_3)$, where $p_i$ refers to the momentum of the pions, $q$ and $\eta$ are the momentum and polarization tensor of the $\omega$ meson, respectively.  This process receive contributions from the $\rho$, $\rho^\prime$ and contact channels, as shown in Figure \ref{ome_3pifig}.
The decay amplitude can be set as 
\begin{equation}
\mathcal{M}_{\omega\rightarrow 3\,\pi} = i\, \epsilon _{\mu \alpha \beta \gamma}\, \eta^{\mu}\, p_1{}^{\alpha }\,p_2{}^{\beta }\,p_3{}^{\gamma }\, \mathcal{A}(m^{2}_{\omega}),
\label{ampf3pi}
\end{equation} 
where $\mathcal{A}(m^{2}_{\omega}) $ is given by
\begin{eqnarray}
\mathcal{A}(m^{2}_{\omega}) &=& 6\,g_{3 \pi} +  2\,g_{\omega \rho \pi }\, g_{\rho \pi \pi }\,\left(D_{\rho^{0}}[s_{12}]+\right. \left. D_{\rho^{+}}[s_{13}]+D_{\rho^{-}}[s_{23}]\right)\nonumber\\
&&\hspace{1 cm}+2\,g_{\omega \rho^{\prime} \pi }\, g_{\rho^{\prime} \pi \pi }\,\left(D_{\rho^{\prime}}[s_{12}]+\right. \left. D_{\rho^{\prime}}[s_{13}]+D_{\rho^{\prime}}[s_{23}]\right),
\label{Apdem}
\end{eqnarray}
and $s_{ij} = p_i+p_j$, $D_{V}[p]=1/(p^2-m_{V}^2+\imath\,m_{V}\,\Gamma _{V})$. The decay widht $\Gamma_V$ is taken as energy dependent for the $\rho$ and $\rho^{\prime}$, while it is taken as a constant for the $\omega$. The factors of 6 and 2 in  $\mathcal{A}(m^{2}_{\omega})$ come from the cyclic permutations and momentum conservation, used to bring the amplitude into the current form. The notation is explicit for the $\rho$ and $\rho^\prime$ contributions.
The decay width is obtained upon integration over the full three-body phase space \cite{pdg}. While for a single decay this procedure faces no major problem, the inclusion in a numerical analysis involving more processes would require a practical approach to speed up. Since we are interested in the couplings (masses and widths are taken at their nominal values), the decay width can be decomposed as a polynomial on the coupling constants as 
\begin{eqnarray}
    \Gamma_{\omega 3\pi} &=& A_1 \, g_{3\pi}^2 + A_2 \, g_{\omega \rho \pi }^2\, g_{\rho \pi \pi }^2 + A_3 \, g_{3\pi}\, g_{\omega \rho \pi }\, g_{\rho \pi \pi } + A_4 \, g_{\omega \rho^{\prime} \pi }^2\, g_{\rho^{\prime} \pi \pi }^2 \nonumber\\
    &&+ A_5 \, g_{\omega \rho^{\prime} \pi }\,g_{3\pi}\, g_{\rho^{\prime} \pi \pi } +  A_6 \, g_{\omega \rho^{\prime} \pi }\, g_{\omega \rho \pi }\,g_{\rho \pi \pi }\,g_{\rho^{\prime} \pi \pi }, 
    \label{w3pexpand}
\end{eqnarray}
where the $A_i$ coefficients can be identified with the corresponding part of the decay width for the couplings involved and are computed following the decay width definition as given by the Particle Data Group (PDG) \cite{pdg}.
The couplings involved in the right-hand side of Eq.~\ref{w3pexpand} are not settled, neither in the theoretical side nor experimentally. Studies on the value of $|g_{\omega \rho' \pi }|$ have found it to lie in the range from 10 to 18 GeV$^{-1}$ \cite{david,prime}. The magnitude for the contact coupling computed in the literature from different approaches is also in a wide range from 29 to 123 GeV$^{-3}$ \cite{rudaz,cesareo,kura,kura1,kay,david}.
We will show that the approach where only the $\rho$ channel is considered requires a large value for the $g_{\omega \rho \pi }$ coupling, compared with the previous estimates considering radiative decays. This result motivates the inclusion of the $\rho(1450)$ and the contact term.

\section{The $e^+e^- \rightarrow \omega \rightarrow 3\,\pi$ cross section}
Now, we proceed to describe the $e^{+}(k_{+})\,e^{-}(k_{-}) \rightarrow \omega(q) \rightarrow \pi^{+}(p_1)\,\pi^{-}(p_2)\,\pi^{0}(p_3)$ cross section. Following the same approach for the hadronic part as in the previous section, but now at an energy of $q^2=(k_{+}+k_{-})^{2}$ instead of $m_\omega^2$, we can write the amplitude for the $\omega$ channel as 
\begin{equation}
\mathcal{M}_{e^{+}\,e^{-}\rightarrow 3\,\pi} = \frac{e}{q^{2}}\frac{m^{2}_{\omega }}{g_{\omega }}\,D_{\omega}(q)\,\mathcal{A}(q^{2})\,\epsilon _{\mu \alpha \beta \gamma }\,p_1{}^{\alpha }p_2{}^{\beta }p_3{}^{\gamma }\,l^{\mu} 
\label{ampee3pi}
\end{equation} 
 where $e$ is the positron electric charge, $l^{\mu} =-i\,e\,\bar{v}(k_{+})\,\gamma^{\mu }\,u(k_{-})$, is the leptonic current. $\mathcal{A}(q^{2})$ has been defined in Eq.~(\ref{Apdem}), but now taken at $q^2$.
Following the same approach as for the $\omega \to 3\,\pi$ decay, we expand the cross section in terms of the coupling constants and coefficients:
\begin{eqnarray}
    \sigma(e^+e^- \to \omega \to 3\,\pi) &=& \frac{1}{g_\omega^2}\,\Big(B_1 \, g_{3\pi}^2 + B_2 \, g_{\omega \rho \pi }^2\,g_{\rho \pi \pi }^2 + B_3 \,  g_{3\pi}\, g_{\omega \rho \pi }\, g_{\rho \pi \pi } + B_4 \, g_{\omega \rho^{\prime} \pi }^2\, g_{\rho^{\prime} \pi \pi }^2 \nonumber\\
    &&+ B_5 \,  g_{\omega \rho^{\prime} \pi }\, g_{3\pi}\, g_{\rho^{\prime} \pi \pi } +  B_6 \,  g_{\omega \rho^{\prime} \pi }\, g_{\omega \rho \pi }\,g_{\rho \pi \pi }\,g_{\rho^{\prime} \pi \pi }\Big), 
\end{eqnarray}  
The $B_i$ are computed only once at each data point energy reported by the experiments, using the kinematical description as given in Ref. \cite{kumar}, and implemented in a Fortran program with the VEGAS \cite{vegas} integration subroutine.
We have considered the data from SND \cite{snd3p}, which uses data from DM2 to extend its range up to 2 GeV. They find evidence of the $\rho \to 3 \pi$ decay mode, with a branching ratio of the order of $10^{-4}$; thus, its smallness justifies our decision of not considering this mode. We consider  the energy range up to around 0.82 GeV, to avoid the $\phi$ contribution.
The CMD2 data \cite{CMD23p} updated the previous measurement \cite{CMD23pold} to include missing contributions in the energy range of 0.76 to 0.821 GeV, in that case they did not perform a spectral analysis. 
For the BABAR \cite{BABAR3p} data, we consider energies below 0.9 GeV to avoid the $\phi$ contribution. They also find that the $\rho \to 3\,\pi$ decay mode, branching ratio is of the order of $10^{-4}$. Preliminary data from BESIII \cite{BES3p} is available in the energy range form 0.7 - 3 GeV, for consistency with the approach we restrict our consideration for this data to energies below 0.8 GeV. Note that these upper values are not the same for all the experiments due to their different energy binning.
We have verified that the particular upper energy value considered in this region makes no effect on the results.

\section{The $e^{+}e^{-}\rightarrow \omega\pi^0\rightarrow \pi^0\pi^0\gamma$ cross section}
\begin{figure}[htb]
\begin{center}
\includegraphics[scale=0.6]{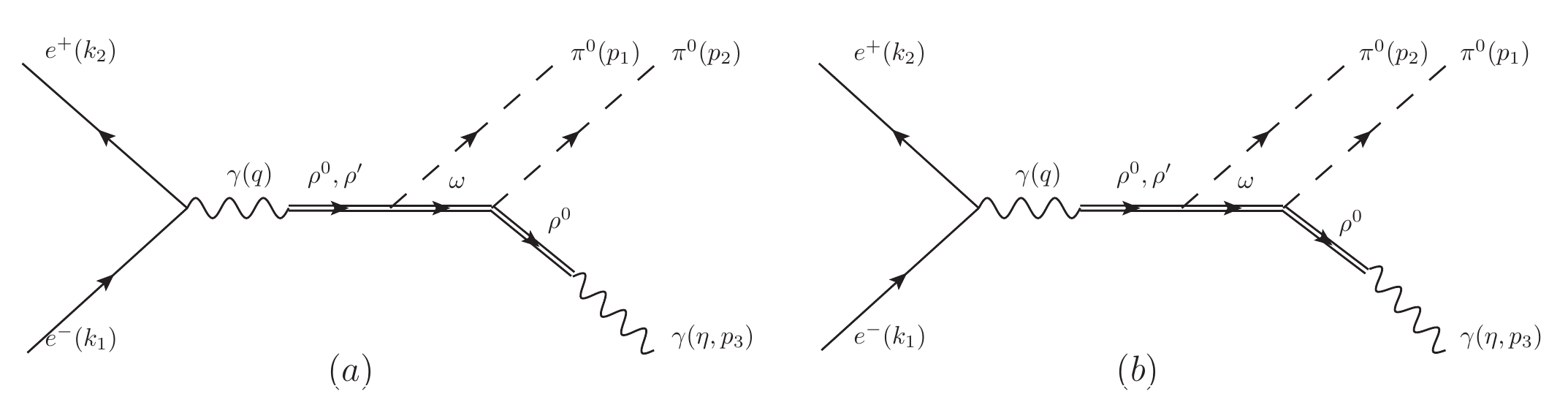}
\end{center}
\caption{The $e^+\,e^- \rightarrow \omega\pi \rightarrow \pi\,\pi\,\gamma$ scattering.}
\label{figppg}
\end{figure}

The notation of momenta for the process is
 $e^{+}(k_{2})\,e^{-}(k_{1})\to \pi^{0}(p_1)\,\pi^{0}(p_2)\,\gamma(\eta^{*},p_3)$, where $\eta^{*}$ represents the polarization tensor of the photon. The process is depicted by the diagrams in Fig.~\ref{figppg}, where both the $\rho$ and $\rho^\prime$ intermediate states are considered.
 Further contributions such as the $\phi$ meson or scalars are not considered at this stage, although they may be relevant for precision observable estimates such as the  muon magnetic dipole moment \cite{davier,jorge}. The amplitude for the diagram of Fig.~\ref{figppg}(a) can be written as: 
\begin{equation}
 \mathcal{M}_{(a)} = \frac{e^{2}}{q^{2}}\,\Big(C_{\rho^{0}}+e^{i\theta}C_{\rho^{\prime}}\Big)\,D_{\omega}(q-p_{1})\,\epsilon_{\mu\sigma\epsilon\lambda}\,q^{\sigma}\,(q-p_1)^{\epsilon}\,{\epsilon_{\alpha\beta\nu}}^\lambda\,(q-p_1)^{\alpha}\,p_3{}^{\beta}\,\eta^{*\nu}\,l^{\mu},
\end{equation}
where the global factors are defined by:
\begin{equation}
 C_{\rho^{0}} = \Big(\frac{g_{\omega\rho\pi}}{g_{\rho}}\Big)^{2}\, m^{2}_{\rho^{0}}\,D_{\rho^{0}}(q), \hspace{0.5 cm} C_{\rho^{\prime}} = \frac{g_{\omega\rho^{\prime}\pi}\,g_{\omega\rho\pi}}{g_{\rho}\,g_{\rho^{\prime}}}\,m^{2}_{\rho^{\prime}}\,D_{\rho^{\prime}}(q),
\end{equation}
with a relative phase $e^{i\theta}$ between both channels.
Note that this amplitude is exactly the same amplitude for Fig.~\ref{figppg}(b) by interchanging $p_1\leftrightarrow p_2$ momenta.

The cross section is set, in terms of the couplings involved and coefficients $C_i$, as:
\begin{equation}
    \sigma(e^+e^-\to 2\pi^0\gamma) = \left(\frac{g_{\omega \rho \pi }}{g_{\rho}}\right)^4 C_1
    +\left(\frac{g_{\omega \rho \pi }}{g_{\rho}}\frac{g_{\omega \rho^{\prime} \pi }}{g_{\rho^{\prime}}}\right)^2 C_2
    +\left(\frac{g_{\omega \rho \pi }^3}{g_{\rho}^3}\frac{g_{\omega \rho^{\prime} \pi }}{g_{\rho^{\prime}}}\right)\Big( Cos(\theta)\, C_3 - Sin(\theta)\,C4\Big). 
\end{equation}  

We have considered the data from three SND Collaboration  measurements \cite{snd2pg00,snd2pg13,snd2pg16}, although the latter \cite{snd2pg16} updated the previous ones, they will be useful to illustrate the behavior of the couplings even in such cases were some corrections are missing. Data from CMD2  Collaboration \cite{cmd22pg} are also available and used in this analysis. We can profit from the corresponding analysis that the experiments carried out, by identifying the parameters region favored from their own fit. In particular, we identify that the relative phase is expected to be large ($\theta = 122\pm 8^0$ is obtained in Ref. \cite{snd2pg16}), and the parameter $A_1\equiv (g_{\omega \rho^{\prime} \pi }/g_{\omega \rho \pi })\,(g_{\rho}/g_{\rho^{\prime}})$ is introduced to describe the process, instead of the individual parameters (do not get confused with $A_1$ coefficient in Eq. (\ref{w3pexpand})). We will consider the latter as a constrain for the individual couplings combination and search for the most favored value. Using the experimental analyses as guidance, we expect it to be around $A_1 \approx 0.2 $.

\section{$\chi^2$ analysis. Results}
\begin{figure}[htb]
\begin{center}
\includegraphics[scale=0.5,angle=-90]{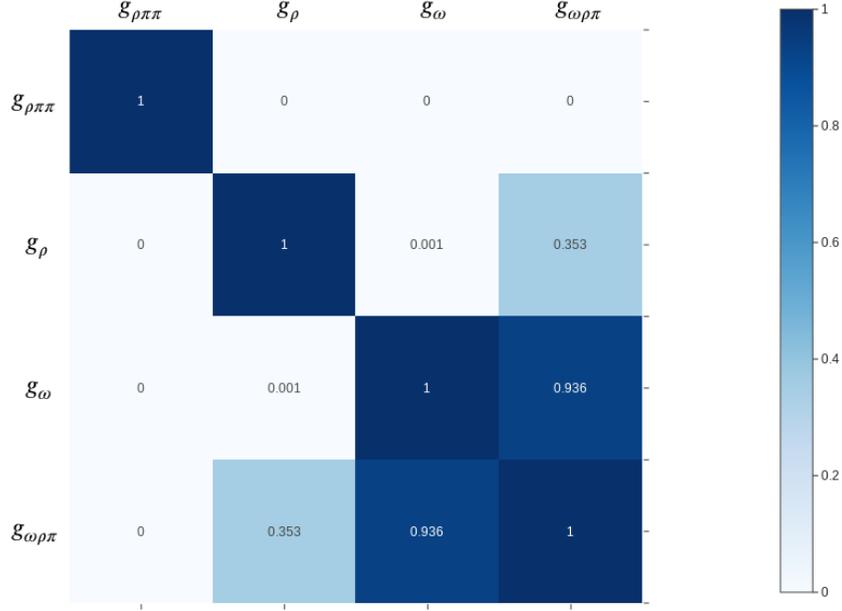}
\end{center}
\caption{Correlation matrix for $g_{\rho \pi \pi }$, $g_{\rho}$, $g_{\omega}$, and $g_{\omega \rho \pi}$ parameters from ten decay modes; see text for details.}
\label{corr10}
\end{figure}

To determine the hadronic couplings of the low-energy mesons and the $\rho^\prime$, from the processes described above, we perform a fit to several datasets available in the literature, using the MINUIT package. The $\chi^2$ function to minimize is defined by
\begin{equation}
\chi^2(\mathbf{\theta})=\sum_{i=1}^N \frac{(y_i-\mu\,(x_i;\mathbf{ \theta}))^2}{E_i^2},
\end{equation} 
where $\mathbf{\theta}=(\theta_1, ..., \theta_N)$ are the parameters to determine; $y_i$ and $E_i$ are the experimental  data and their corresponding uncertainties. $\mu(x_i;\mathbf{\theta})$ are the theoretical estimates for the corresponding parameters.
In a first step, we determine the parameters of the model involving the light mesons, from ten decay modes which are insensitive to the $\rho'(1450)$, namely, $\rho \to \pi\,\pi$ neutral and charged modes;  $\rho^0 \to e^+\,e^-, \ \mu^+\,\mu^-$, $\omega \to e^+\,e^-, \ \mu^+\,\mu^-$, $\omega \to \pi^0\,\gamma$, $\rho \to \pi\,\gamma$ neutral and charged modes; and $\pi^0 \to \gamma\,\gamma$, using the experimental information of each decay width and the mass of the particles involved, as listed in the PDG \cite{pdg}. These processes involve four parameters: $g_{\rho}$, $g_{\rho \pi \pi }$, $g_{\omega}$ and $g_{\omega\rho\pi}$. In Table \ref{fit10}, we show the result of the fit. The value of the minimization function per degree of freedom ($dof$) is $\chi^2/dof=0.32$. The correlation between parameters is shown in Fig. \ref{corr10} as a heat map. The large correlation between $g_{\omega}$ and $g_{\omega\rho\pi}$ is due to the fact that this set of observables involve both, the $\omega-\rho-\pi$ interaction and the $\omega$-photon transition combined.
\begin{table}[htb]
\begin{center}
\begin{tabular}{lcc}
     \hline
     \hline
     Parameter & &Value \\ 
     \hline 
    $g_{\rho \pi \pi }$ & &5.949 $\pm$ 0.054  \\  
    $g_{\rho}$  & &4.962 $\pm$ 0.066  \\ 
    $g_{\omega}$  & &17.038 $\pm$ 0.603  \\ 
    $g_{\omega\rho\pi}$ (GeV$^{-1})$ & & 11.575 $\pm$ 0.438  \\ 
    \hline
    \hline
\end{tabular}
\end{center}
\caption{Parameters from the fit to ten decay modes as described in the text.}
\label{fit10}
\end{table}

\begin{figure}[htb]
\begin{center}
\includegraphics[scale=0.6,angle=-90]{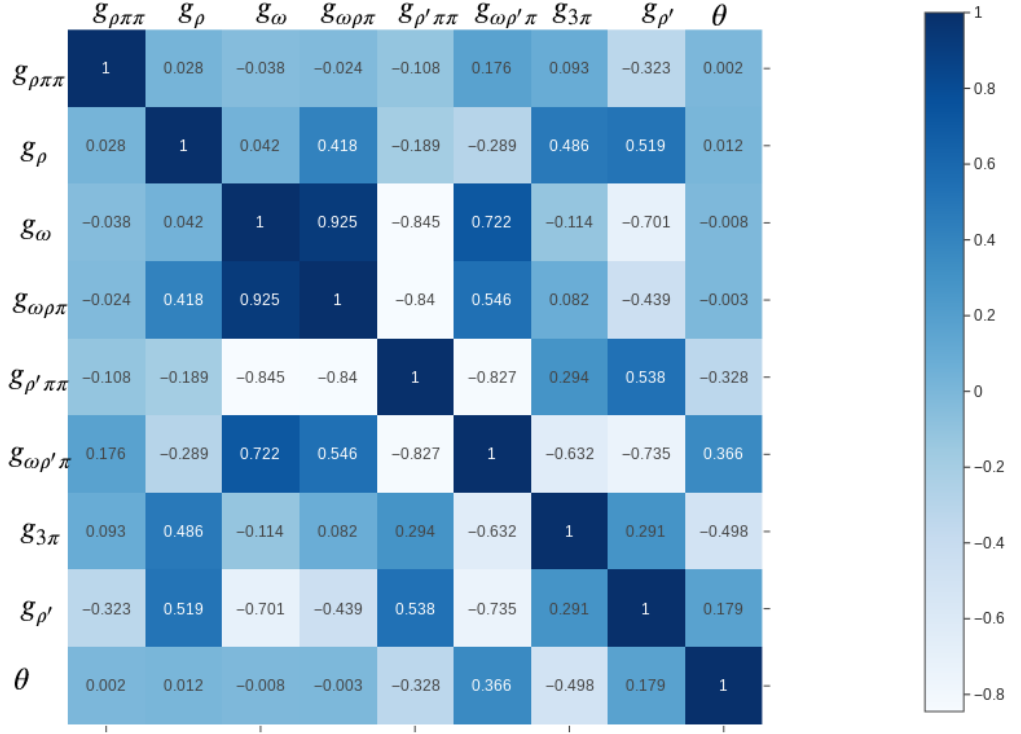}
\end{center}
\caption{Correlation matrix for the couplings considering 11 decay modes and data for $e^+\,e^- \rightarrow \pi^0\,\pi^0\,\gamma$ cross section. See the text for details.}
\label{heat2}
\end{figure}

\begin{figure}[htb]
\begin{center}
\includegraphics[scale=0.6,angle=-90]{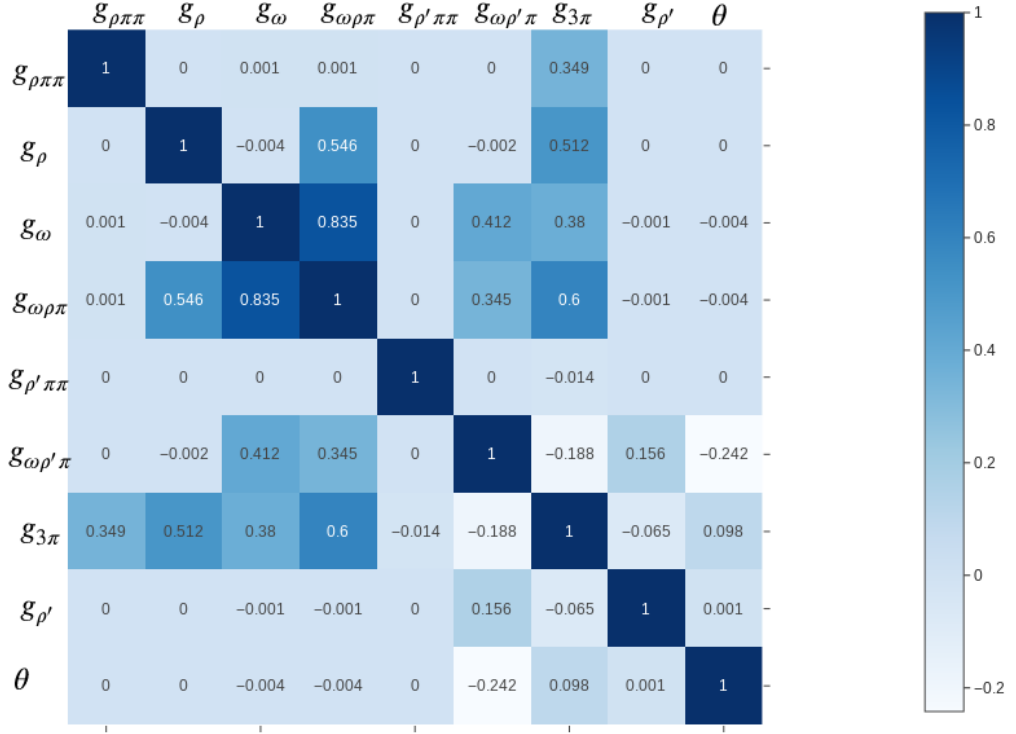}
\end{center}
\caption{Correlation matrix for the couplings considering 11 decay modes and data for $e^+\,e^- \rightarrow \pi^0\,\pi^0\,\gamma$ and  $e^+\,e^- \rightarrow \omega \rightarrow 3\,\pi$ cross sections. See the text for details.}
\label{heat3}
\end{figure}

\begin{figure}[htb]
\vspace*{.5cm}
\begin{center}
\includegraphics[scale=0.55]{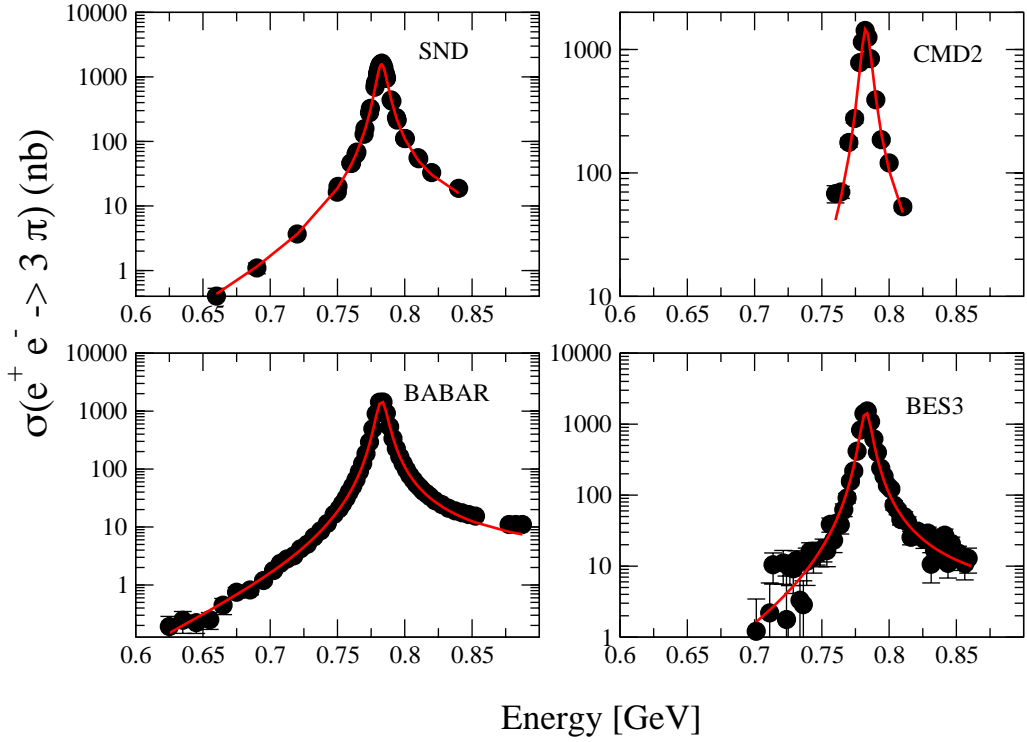}
\end{center}
\caption{ $e^+e^- \rightarrow \omega \rightarrow 3\,\pi$ cross section data (symbols) from SND \cite{snd3p}, CMD2 \cite{CMD23p}, BABAR \cite{BABAR3p}, and BESIII \cite{BES3p} and the corresponding results (solid line) using the parameters from the global analysis, Table \ref{fit2pg3p}.}
\label{x3fit}
\end{figure}

\begin{figure}[htb]
\vspace*{.5cm}
\begin{center}
\includegraphics[scale=0.55]{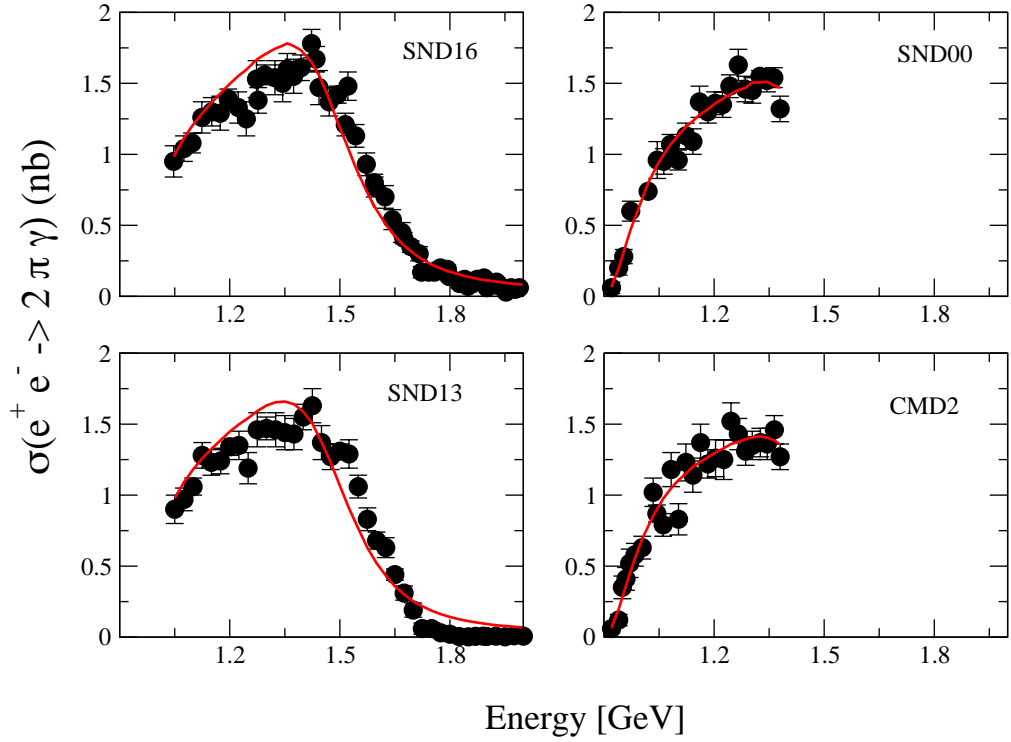}
\end{center}
\caption{ $e^+\,e^- \rightarrow \pi^0\,\pi^0\,\gamma$ cross section data (symbols) from SND (SND00 \cite{snd2pg00} SND13 \cite{snd2pg13} and SND16 \cite{snd2pg16}  and CMD2 \cite{cmd22pg} and the corresponding results (solid line) using the parameters from the analysis, Table \ref{fit2pg}.}
\label{x2fit}
\end{figure}

\begin{figure}[htb]
\begin{center}
\includegraphics[scale=0.6]{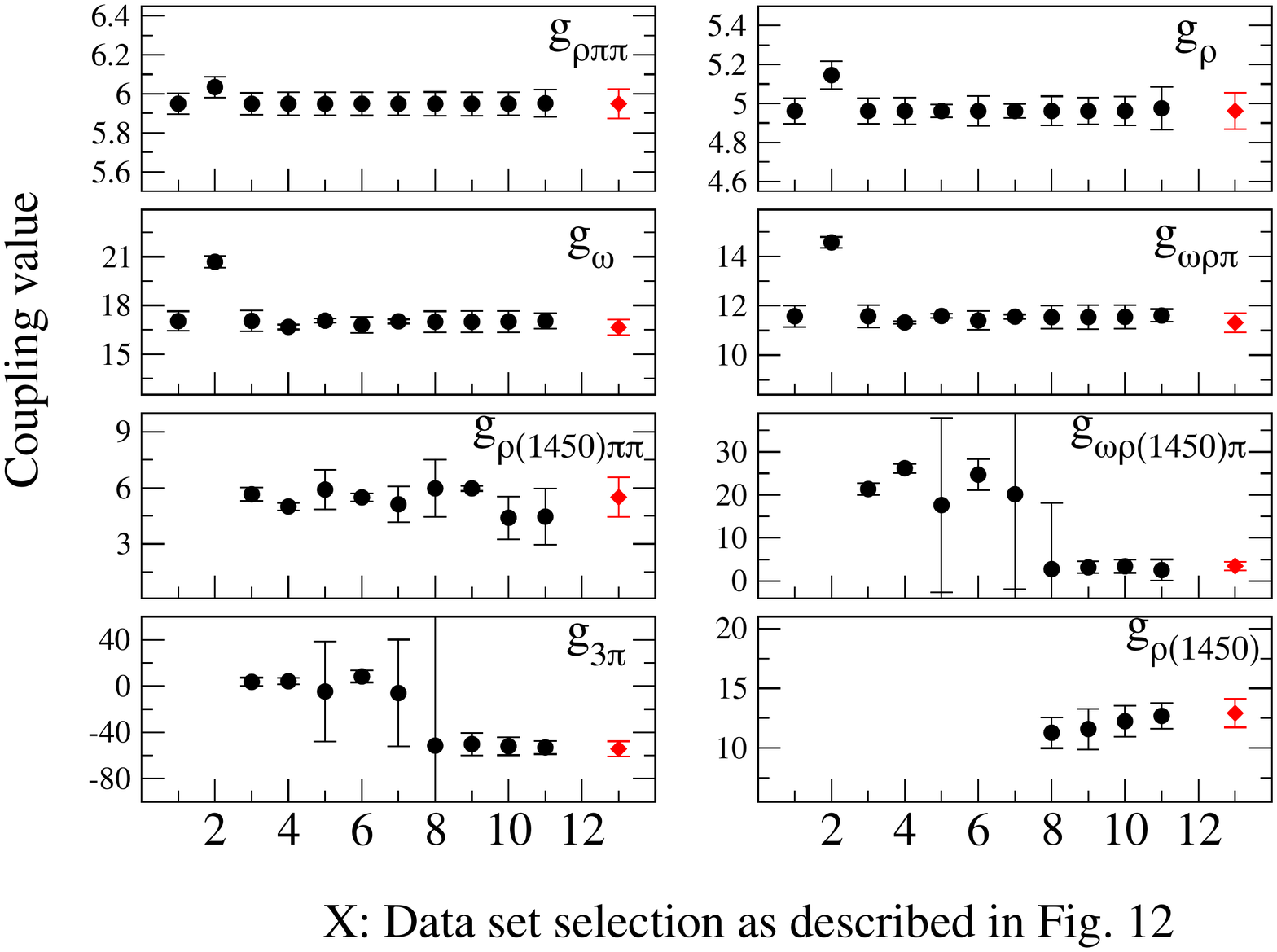}
\end{center}
\caption{All the coupling parameters involved in the analysis, with their corresponding errors as a function of the dataset considered for the fit, to exhibit the variations as a function of them. See Fig. \ref{labelx} for the labeling on the $x$ axis.} 
\label{couplings}
\end{figure}

\begin{figure}[htb]
\begin{center}
\includegraphics[scale=0.5]{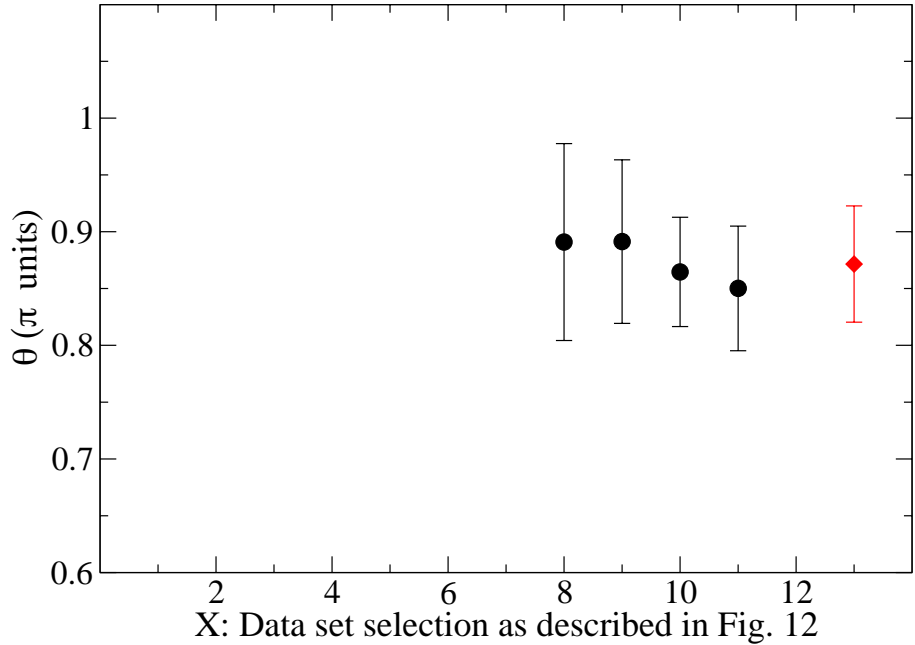}
\end{center}
\caption{Relative phase parameter $\theta$ as a function of the dataset considered for the fit, to exhibit the variations as a function of them. See Fig. \ref{labelx} for the labeling on the $x$ axis.} 
\label{angle}
\end{figure}

\begin{figure}[htb]
\begin{center}
\includegraphics[scale=0.5]{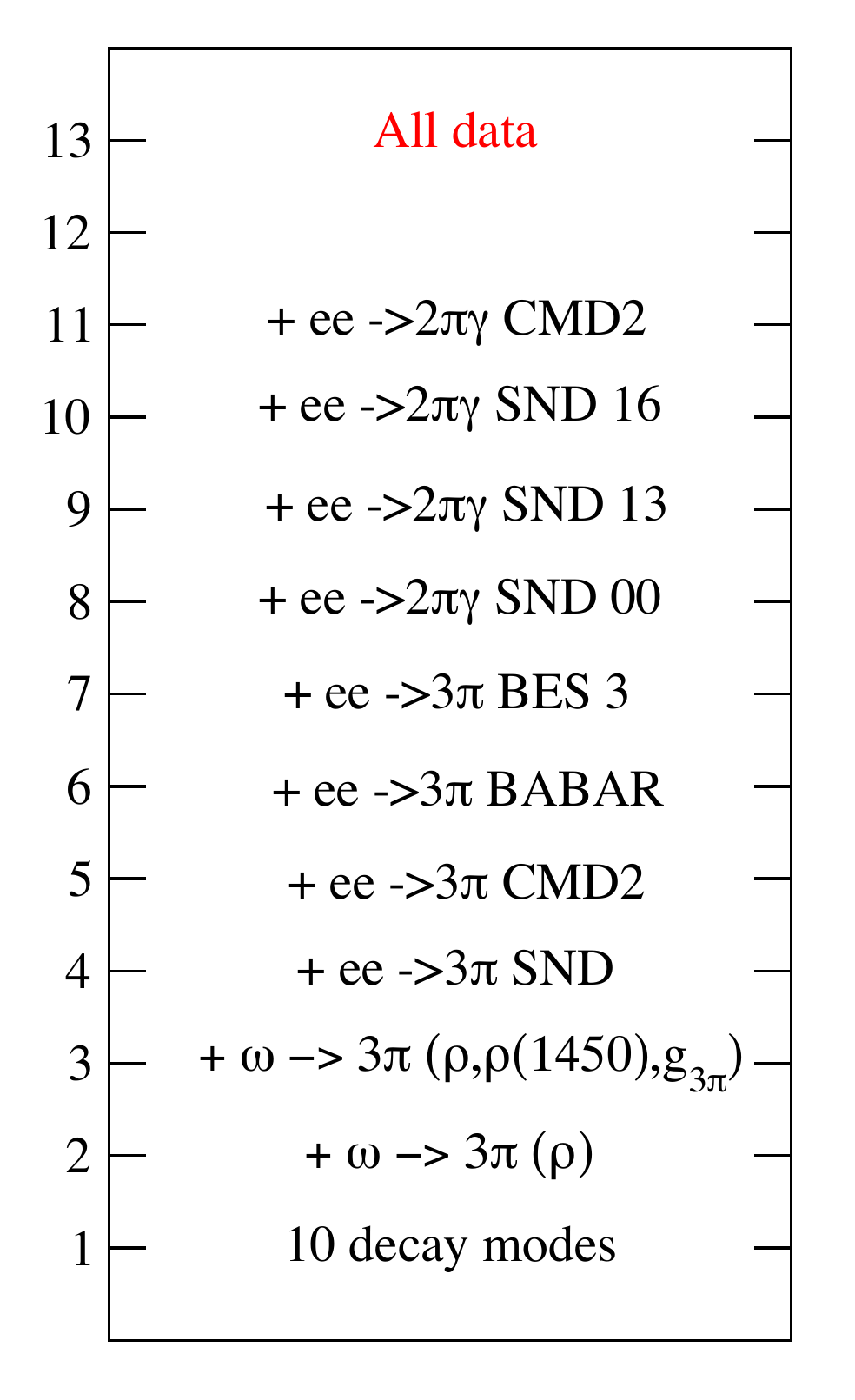}
\end{center}
\caption{Label for the $x$ axis in Figs.~\ref{couplings} and \ref{angle}. In 2 and 3, the (+) symbol means in addition to the case at 1 (that is 10 decay modes). In 4 to 11, the (+) symbol means in addition to the case at 3 (That is 11 decay modes). {\it All data}, means considering all the previous cross section data and case at 3.} 
\label{labelx}
\end{figure}
 
Then, we include the $\omega \rightarrow 3\,\pi$ decay mode to exhibit the strong modification of the $g_{\omega\rho\pi}$ parameter previously obtained, which becomes $g_{\omega\rho\pi}=14.57 \pm 0.22$ and a $\chi^2/dof>>1$, signaling the inconsistency and therefore the need to extend the description by incorporating the $\rho(1450)$ and a contact term as prescribed by the WZW anomaly. Upon the inclusion of these contributions we obtain $g_{\omega\rho\pi}=11.576 \pm 0.463$, in accordance with previous results. Hereafter this is the way to describe the $\omega$ decay, and denote this set of data as the {\it 11 decay modes}. In a second step, we incorporate the data from the $e^+\,e^- \rightarrow 3\,\pi$ cross section (as measured by SND \cite{snd3p}, CMD2 \cite{CMD23p}, BABAR \cite{BABAR3p} and BESIII \cite{BES3p}) and the $e^+\,e^- \rightarrow \pi^0\,\pi^0\,\gamma$ (as measured by SND \cite{snd2pg00,snd2pg13,snd2pg16} and CDM2 \cite{cmd22pg}) to further restrict the $\rho(1450)$ parameters validity region. Global restrictions from other measurements, as the mentioned $A_1$ and the upper bound for the $g_{\rho^\prime \pi \pi }$ parameter, are incorporated by setting a consistent region for the search of the parameters in the minimization process.
In particular, we obtain $A_1=0.125 \pm 0.05$.
Table \ref{fit2pg} shows the parameters value when considering the 11 decay modes plus the experimental data for $e^+\,e^- \rightarrow \pi^0\,\pi^0\,\gamma$ cross section, and Table \ref{fit2pg3p} corresponds to the results when adding $e^+\,e^- \rightarrow 3\,\pi$ cross section data. The corresponding correlation  matrix are shown as heat maps in Figs. \ref{heat2} and \ref{heat3}, respectively. In Fig.~\ref{x3fit}, we show the $e^{+}e^{-}\to\pi\omega\to3\,\pi$ cross section data from several measurements (symbols) and the result from the analysis (solid line) considering the 11 decay modes and all the cross section data, Table \ref{fit2pg3p}. In Fig.~\ref{x2fit}, we show the $e^{+}e^{-}\to\pi^{0}\pi^{0}\gamma$ cross section data from several measurements (symbols) and the result from the analysis (solid line) considering the 11 decay modes and the cross section data for $e^{+}e^{-}\to\pi^{0}\pi^{0}\gamma$, Table \ref{fit2pg}.\\
To summarize the results, in Figs. \ref{couplings} and \ref{angle} we have plotted the values of the individual parameters as a function of the dataset considered for the minimization ($x$ axis).
$x=1$ considers the ten decay modes dataset mentioned above; notice that only the four parameters involved exhibit a symbol. 
$x=2$ considers the dataset of the ten decay modes plus the $\omega \to  3\pi$ decay, described considering only the $\rho$ as intermediate state.
$x=3$ considers the dataset of the 11 decay modes as described before. The parameters for $x=$4,...,11 are the corresponding results when adding individual cross section data to the 11 decay modes. Namely,
$e^+\,e^- \rightarrow 3\,\pi$ from SND ($x=4$) \cite{snd3p},
 CMD2 ($x=5$) \cite{CMD23p}, BABAR ($x=6$) \cite{BABAR3p},and BESIII ($x=7$) \cite{BES3p}, and the $e^+\,e^- \rightarrow \pi^0\,\pi^0\,\gamma$ from SND ($x=8$) \cite{snd2pg00}, ($x=9$) \cite{snd2pg13} ,  ($x=10$) \cite{snd2pg16} and CDM2  ($x=11$) \cite{cmd22pg} . $x=12$ is left empty and $x=13$ corresponds to the case including all the experimental data (we use red color symbols to make a further distinction with respect to the other cases).
For the sake of clarity, in Fig. \ref{labelx}, we show the description of them, corresponding to the $x$-axis labeling of the previous figures. Missing parameter data in any of these $x$ values means that the dataset has no dependence on it.

\begin{table}[htb]
\begin{center}
\begin{tabular}{lcc}
     \hline
     \hline
     Parameter & &Value \\ 
     \hline
     \hline
    $g_{\rho \pi \pi }$ & &5.948    $\pm$   0.067 \\
    $g_{\rho}$  & & 4.962    $\pm$  0.082\\
    $g_{\omega}$  & &  16.907   $\pm$    0.663 \\
    $g_{\omega\rho\pi}$ (GeV$^{-1})$ & &11.486  $\pm$ 0.495 \\ 
    $g_{\rho^\prime \pi \pi }$ &  & 4.51    $\pm$    1.037\\  
    $g_{\omega\rho^\prime\pi}$ (GeV$^{-1})$ & & 3.136  $\pm$ 1.77 \\ 
    $g_{3 \pi }$ (GeV$^{-3}$)& & -53.612 $\pm$   6.893\\
    $g_{\rho^\prime}$  &  & 12.472      $\pm$  1.244  \\ 
 $\theta$ (in $\pi$ units)&  & 0.87    $\pm$   0.045\\
    \hline
    \hline
\end{tabular}
\end{center}
\caption{Parameters from the fit to 11 decay modes and cross section data for $e^+\,e^- \to \pi^0\,\pi^0\,\gamma$.}
\label{fit2pg}
\end{table}
\begin{table}[htb]
\begin{center}
\begin{tabular}{lcc}
     \hline
     \hline
     Parameter & &Value \\ \hline 
    $g_{\rho \pi \pi }$ & &5.949    $\pm$   0.076  \\
 
    $g_{\rho}$  &  &4.962   $\pm$    0.093\\
    $g_{\omega}$  &  &16.652   $\pm$    0.473\\
    $g_{\omega\rho\pi}$ (GeV$^{-1})$ & &11.314 $\pm$ 0.383\\ 
    $g_{\rho^\prime \pi \pi }$ & &5.5    $\pm$   1.06 \\   
    $g_{\omega\rho^\prime\pi}$ (GeV$^{-1})$ & &3.477 $\pm$ 0.963\\ 
    $g_{3 \pi }$ (GeV$^{-3}$)& &-54.338  $\pm$ 6.674\\
    $g_{\rho^\prime}$  &  &12.918 $\pm$   1.191 \\ 
 $\theta$ (in $\pi$ units)& &0.872 $\pm$  0.051\\
\hline
\hline
\end{tabular}
\end{center}
\caption{Parameters from the fit to the 11 decay modes and all the cross section data.}
\label{fit2pg3p}
\end{table}
 
\subsection{The $e^+\,e^- \rightarrow \pi\,\omega \rightarrow 4\,\pi$ cross section}
\begin{figure}[htb]
\begin{center}
\includegraphics[scale=0.55]{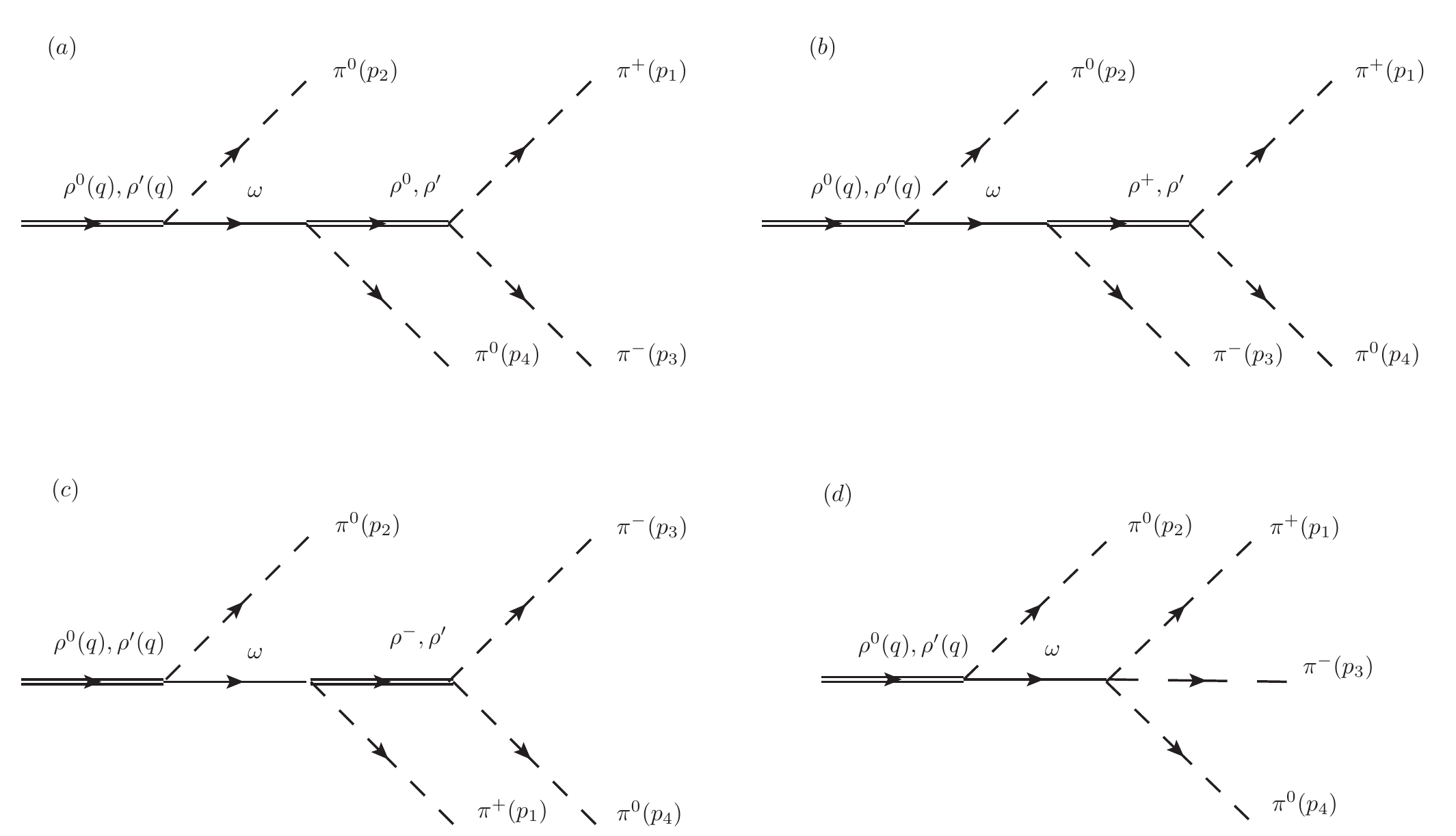}
\end{center}
\caption{The $e^+\,e^- \rightarrow \pi\,\omega \rightarrow 4\,\pi$ cross section due to the $\omega$ channel.} 
\label{4pifig}
\end{figure}

As an application of the results, we compute the $e^+\,e^- \rightarrow 4\,\pi$ cross section for the so-called omega channel and compare with the data reported by BABAR \cite{babar4p}, we do not consider the recent measurement from SND \cite{snd4p} since explicit data is not provided.
This process has been considered in previous studies  to test models' viability to account for the observed data, to study isospin symmetry breaking effects as compared with the analog in tau decays \cite{Ecker4pi,Czyz4pi}, and to determine the magnetic dipole moment of the $\rho$ \cite{davidmdm}. Here, we do not fit the data but use the parameters found as listed in Tables \ref{fit2pg} and \ref{fit2pg3p} to describe it.
The $\omega$ channel is depicted in Fig. \ref{4pifig}. Let us set the notation for the momenta of the process as follows: $e^+(k_+)\,e^-(k_-) \rightarrow \pi^{+}(p_1)\,\pi^{0}(p_2)\,\pi^{-}(p_3)\,\pi^{0}(p_4)$. Then, we can write the amplitude as
\begin{equation}
\mathcal{M}_{e^{+}e^{-}\rightarrow 4\,\pi} = \frac{e}{q^{2}} \,\Big(G_{\rho}+e^{i\theta}G_{\rho^{\prime}}\Big)\,D_{\omega}(q-p_4)\,\mathcal{A}((q-p_4)^2)\,\epsilon_{\sigma\alpha\eta\beta}\,{\epsilon_{\mu\gamma \chi}}^\sigma \,q^{\gamma}\,p_1{}^{\alpha}\,p_2{}^{\eta}\,p_3{}^{\beta}\,p_4{}^{\chi}\,l^{\mu}\, ,
\label{ampee4pi}
\end{equation}
where
\begin{equation}
G_{\rho} = \frac{g_{\omega\rho\pi}}{g_{\rho}}m^{2}_{\rho^{0}}D_{\rho^{0}}(q),\hspace{0.5 cm}G_{\rho^{\prime}} = \frac{g_{\omega\rho^{\prime}\pi}}{g_{\rho^{\prime}}}m^{2}_{\rho^{\prime}}D_{\rho^{\prime}}(q).    
\end{equation}
The Bose-Einstein symmetry, applied to the neutral pions, leads to an additional contribution by the momentum exchange of the neutral pions in all diagrams. The corresponding amplitude is similar to Eq.(\ref{ampee4pi}) by exchanging $p_4\leftrightarrow p_2$.

\begin{figure}[htb]
\begin{center}
\includegraphics[scale=0.55]{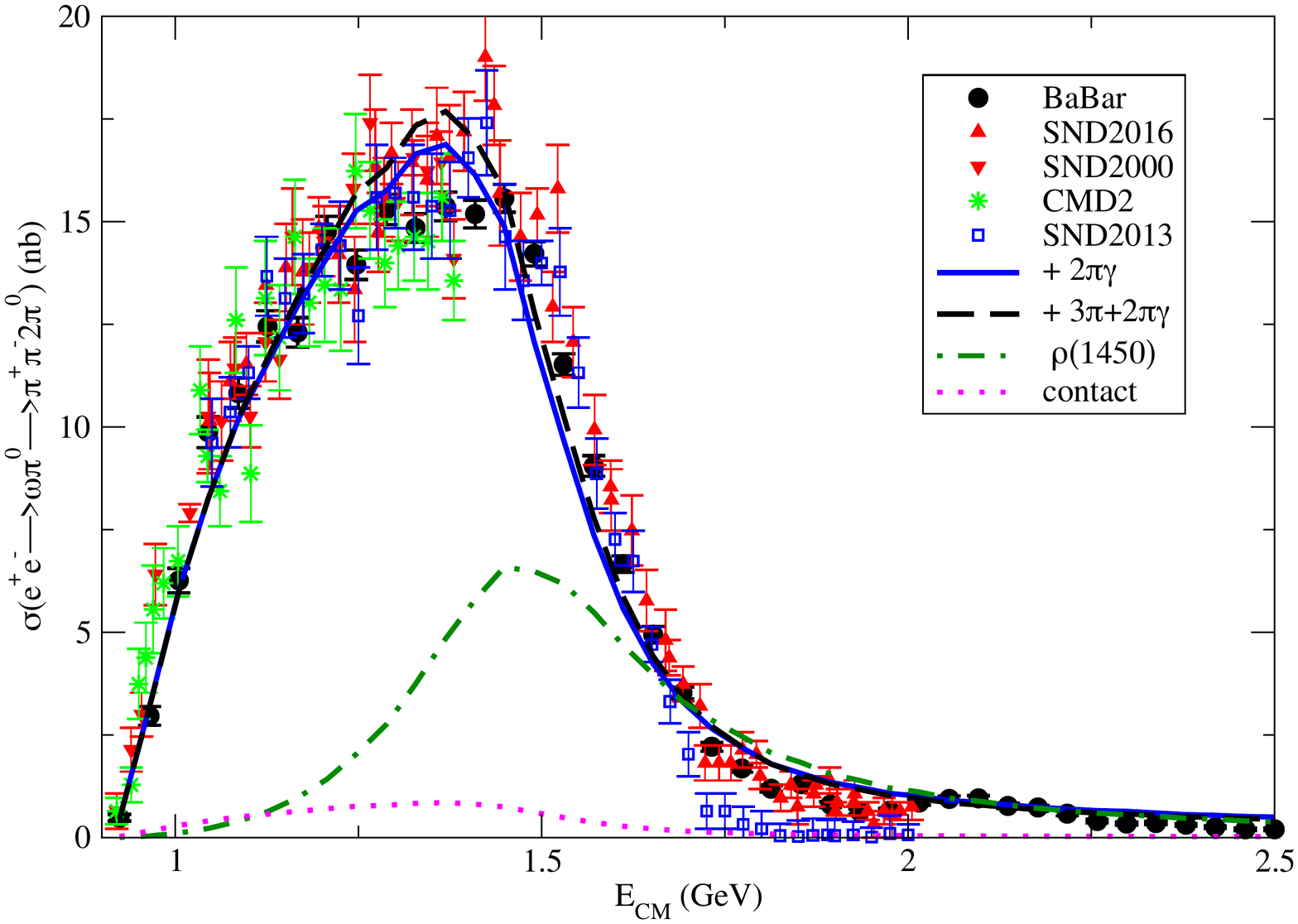}
\end{center}
\caption{The $e^+\,e^- \rightarrow \pi\,\omega \rightarrow 4\pi$ cross section driven by the $\omega$ channel. Symbols  correspond to experimental data. Lines correspond to the evaluation in the model considering the parameters determination from the different datasets, Table \ref{fit2pg} (dashed line) and \ref{fit2pg3p} (solid line). The $\rho^\prime$ (dot-dashed line) and contact contribution (dotted line) are also shown.}
\label{4piout}
\end{figure}

In Fig. \ref{4piout}, we plot the cross section for the parameters values on Table \ref{fit2pg} (dashed line) and  Table \ref{fit2pg3p} (solid line). Experimental data from BABAR \cite{babar4p} are shown in circle symbols. Scaled data from SND \cite{snd2pg00,snd2pg13,snd2pg16} and CMD2 \cite{cmd22pg} obtained from the $e^{+}\,e^{-}\rightarrow \pi\,\pi\,\gamma$ cross section measurements are also displayed (see detailed description inside the figure). We observe that there is a proper description of the data using either parameters data set. The $\rho^\prime$ contribution is shown (dashed line) to illustrate its relevance, a nontrivial role of the interference is important (and therefore the $\theta$ phase) to properly account for the data. The nonresonant contribution ($g_{3\pi}$) coming from the $\omega$ decay is also shown (dotted line).

\section{Discussion}
We have explored the role of the $\rho^\prime$ in low-energy observables by performing an analysis of a set of decay modes and cross sections. 
In a first step, we determined the parameters of the model involving the light mesons, from ten decay modes which are insensitive to the $\rho^\prime$. This provided the ground for the expected region for these parameters, and the reference for the evolution of their behavior, as observables sensitive to the $\rho^\prime$ were added. The incorporation of the $\omega \rightarrow 3\,\pi$ decay, considering only the $\rho$ as intermediate state, induced a strong departure of the $g_{\rho\omega\pi}$ coupling from the previously obtained values. The other couplings involved, namely, $g_\rho$, $g_{\rho\pi\pi}$ and $g_\omega$, also reflected this tension (jump on these parameters in Fig. \ref{couplings} at $x$ axis value 2). 
Extending the description by incorporating the $\rho^\prime$ and a contact term as prescribed by the WZW anomaly  brought this parameter to peace with the previous data results (in Fig. \ref{couplings}, $x$-axis value 3). Upon the incorporation of the data from the $e^+\,e^- \rightarrow 3\,\pi$  (as measured by SND, CMD2, BABAR and BESIII),
and $e^+\,e^- \rightarrow \pi^0 \,\pi^0 \,\gamma$  (as measured by SND and CDM2) cross section data, we were able to further restrict the $\rho^\prime$ parameters validity region. The analysis exhibited the sensitivity to the relative phase and the $\rho^\prime$ parameters. The restriction on the relation between parameters, encoded in $A_1$, was very useful to bring the parameters within the physically expected region. 
The behavior of the $g_{\omega\rho^\prime\pi}$ and $g_{3\pi}$ parameters reflected a process dependence as they favored different values for  $e^+\,e^- \rightarrow 3\,\pi$ compared to $e^+\,e^- \rightarrow \pi^0 \,\pi^0 \,\gamma$ cross section. This may be due to the different experimental precision and energy region scanned by each process. In addition, since the $g_{3\pi}$ is a nonresonant contribution, it can be affected by the background subtraction procedure followed by the experiments. All over, we observed that the behavior of the $g_{\rho\omega\pi}$ coupling was very stable, upon the inclusion of the $\rho^\prime$ and the contact term, favoring a value of $g_{\rho\omega\pi}= 11.314 \pm 0.383$ GeV$^{-1}$ for all the experimental data. This is important as this parameter has implications on other observables related with precision physics \cite{Flores-Baez:2006yiq,Flores-Tlalpa:2005msx,Davier:2010fmf,Miranda}. 
As an application of the results, we computed the $e^+\,e^- \rightarrow 4\,\pi$ cross section for the so-called omega channel, considering the parameters found, and compared with the data measured by BABAR. The data was properly described, exhibiting the importance of the $\rho^\prime$ and the contact term. This channel plays an important role to extract further information from the total $4\pi$ process \cite{davidmdm}, as it provides the dominant contribution at low energies.\\
This analysis has exhibited the importance of the $\rho^\prime$ meson and provided a reliable region for its parameters. This is not an exhaustive analysis since more processes such as $\tau$ decays were not included, but provides the ground to extend them (as we plan to do) while already points out to definite regions for the parameters, that can be useful to describe other processes.\\ 

\begin{acknowledgments}
We acknowledge the support of CONACyT, Mexico Grants 332429 (M. S) and 711019 (A.R.) and the support of DGAPA-PAPIIT UNAM, under grant no. IN110622, PRIDIF IFUNAM fellowship (A.R.). We thank Dr. Roelof Bijker for reading the manuscript and Dr. Pablo Roig and Dr. I. Heredia de la Cruz for very useful comments.
\end{acknowledgments}

\end{document}